Perspective

# Removing muscle artifacts from EEG data of people with cognitive impairment using high order statistic methods

Kalogiannis Grigorios[1] M.Sc, PhD Cand, Chassapis George[1] M.Sc, PhD, Tsolaki Magda[2,3] MD, PhD.

1. School of Electrical and Computer Engineering, Faculty of Engineering, Aristotle University of Thessaloniki (AUTh), Greece
2. Greek Association of Alzheimer's Disease and Related Disorders, Thessaloniki, Greece (GAADRD)
3. Department of Neurology, School of Medicine, Aristotle University of Thessaloniki (AUTh), Greece

**Corresponding author:**
Kalogiannis Grigorios, PhD Candidate, Computer Systems and Architecture Laboratory (CSAL), School of Electrical and Computer Engineering, Faculty of Engineering, Aristotle University of Thessaloniki, Tel: +302310995922, email: gkalogiannis@ece.auth.gr

## Abstract

**Objective**: Often, people with Subjective Cognitive Impairment (SCI), Mild Cognitive Impairment (MCI) and dementia are underwent to Electroencephalography (EEG) in order to evaluate through biological indexes the functional connectivity between brain regions and activation areas during cognitive performance. EEG recordings are frequently contaminated by muscle artifacts, which obscure and complicate their interpretation. These muscle artifacts are particularly difficult to be removed from the EEG in order the latter to be used for further clinical evaluation. In this paper, we proposed a new approach in removing muscle artifacts from EEG data using a method that combines second and high order statistical information. **Subjects and Methods**: In the proposed system the muscle artifacts of the EEG signal are removed by using the Independent Vector Analysis (IVA). The latter was formulated as a general joint Blind Source Separation (BSS) method that uses both second-order and higher order statistical information and thus takes advantage of both Independent Component Analysis (ICA) and Canonical Correlation Analysis (CCA). Diagonalization methods for IVA in the proposed system were reworked based on SCHUR decomposition offering a faster second order blind identification algorithm that can be used on time demanding applications. **Results**: The proposed method is evaluated in both simulated and real EEG data. To quantitatively examine the performance of the new method, two objective measures were adopted. The first measure is the Root Mean Square Error (RMSE) while the second is the Signal-to-noise-ratio (SNR). **Conclusion**: The proposed method overcomes with the need of removing muscle artifacts on both realistic simulated EEG data and brain activity from people with cognitive impairment.

## Introduction

The EEG is a recording of the electrical activity of the brain and reflects the summation of postsynaptic potentials of groups of cortical neurons arranged perpendicular to the scalp. The EEG is frequently contaminated by electrophysiological potentials associated with muscle contraction due to biting, chewing and frowning. These muscle artifacts, obscure the EEG and complicate the interpretation of the EEG or even make the interpretation unfeasible. Hence, there is a clear need to remove these artifacts from the EEG. A simple technique is to uses low-pass filters. However, as the frequency spectrum of the muscle artifacts projects with the frequency spectrums of that of brain signals, frequency filters not only remove the muscle artifacts but also necessary EEG information. Regression methods, investigated for eye movement artifact removal are not adapted for use, because no reference channel is available [1,2].

A more recently solution to this problem is the Independent Component Analysis (ICA) which separates the EEG into statistical independent components [3,4]. This method was already successfully applied to ocular artifact removal [5]. However, cross-talk can be observed when the separation of brain and muscle activity is considered. Furthermore, when using the ICA, identification of the components containing artifacts such as muscle activity, is not obvious, thus further user attention is needed [6]. ICA techniques that try to solve this problem, such as constrained ICA (cICA), cannot be applied for muscle artifact removal since this method locates only that component that is most common to a specific reference signal [7].

The ICA is a standard Blind Source Separation (BSS) method, which works under the assumption that sources are mutually independent, and that the mixing procedure is linear and instantaneous. Applications of BSS techniques can include speech enhancement, robust speech recognition, analyzing EEG and fMRI signals, feature extraction, image de-noising, etc. The most common ICA algorithms used in EEG data analysis are Infomax ICA [8,9], SOBI [10], and FastICA [11]. However, signals are often mixed in a convoluted manner. One common way to extend the instantaneous ICA to the convoluted model is the frequency domain blind source separation (FDBSS) approach. In FDBSS, observed signals are transformed to time-frequency (T-F) domain using short time Fourier transform (STFT). Although FDBSS has many advantages, it suffers from the well-known "permutation problem" that occurred when separated data must be aligned to make sure that each output signal only contains data from the same source [12,13].

The Independent Vector Analysis (IVA) was developed as an extension of ICA. Sources in the IVA model are considered as vectors instead of scalars. IVA utilizes not only the statistical independency among different sources, but also the statistical inner dependency of each source vector. The largest advantage of IVA is that the permutation problem is automatically avoided, and therefore there is no need for a post processing step after ICA for source alignment [14].

In the proposed system the muscle artifacts of the EEG signal are removed by using the IVA. This proposed method uses both second-order and higher order statistical information and thus takes advantage of both ICA and Canonical Correlation Analysis (CCA). During process, we assume that a linear mixing model exists in each dimension separately, and that the latent sources are seperated from others. In contrast to the ICA method, the sources can be random vectors, and therefore the elements of the later are closely related. In IVA, the goal is mixture identification or signal separation for a collection of disjoint but coupled data sets.

## Subjects and Methods

The proposed system is illustrated in Figure 1. EEG datasets are considered as contaminated signals that must be pre-processed. Primary, a signal filtering component was design for performing a band pass filtering. The produced filtered data is then leaded to the IVA-BSS component which will separate muscle artifacts from the signal. Pure EEG signals are then evaluated using the performance evaluation component. Filter implementation, IVA-BSS analysis and performance evaluation were developed using Matlab V2013a.

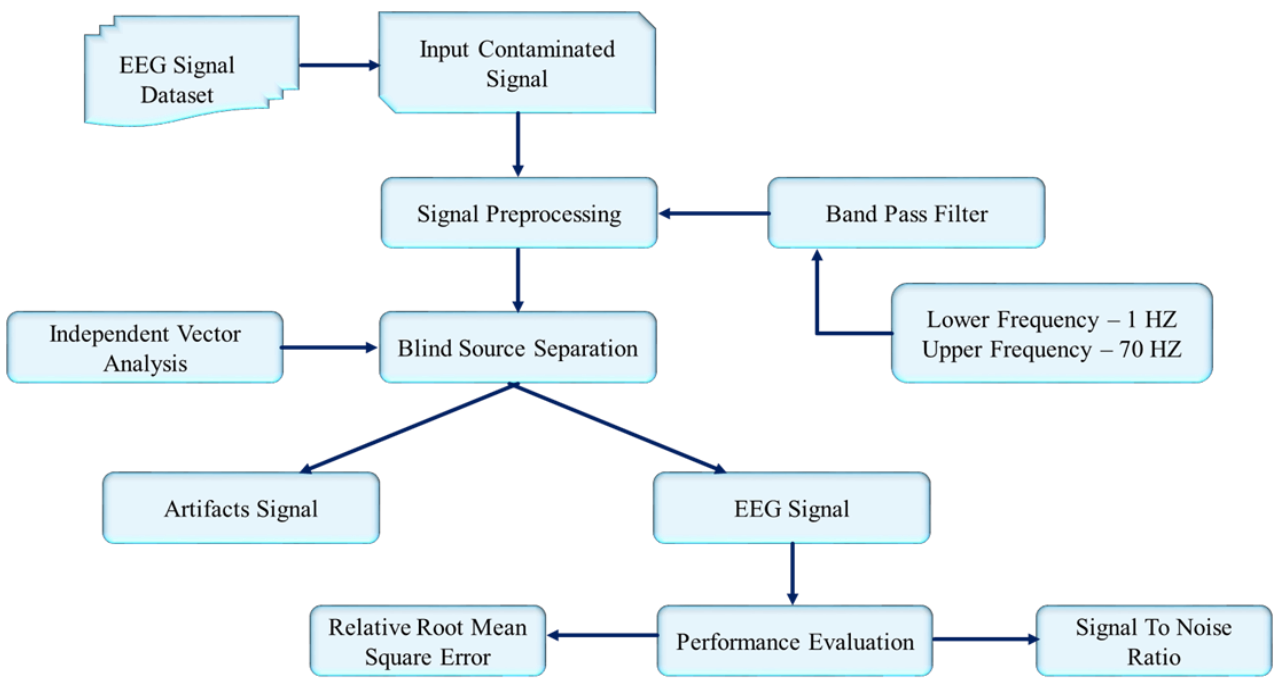

**Figure 1.** The proposed system using the IVA as BSS technique

### EEG Signal Datasets

We first validate and evaluate our methods on several different realistic simulated data. The method for generating realistic simulated data was proposed by Xun [15]. We then apply them to real EEG recordings recruited from the Day Care Center of the Greek Association of Alzheimer Disease and Related Disorders "Saint ioannis", Thessaloniki. These real EEGs were acquired by using the Nihon Kohden EEG-1100C V01.00 system. The sampling frequency was at 500 Hz while 19 electrodes (FP1, FP2, F3, F4, C3, C4, P3, P4, O1, O2, F7, F8, T3, T4, T5, T6, Fz, Cz,

Pz) were placed according to the 10-20 system. Additional validation and evaluation was performed on these real EEG data.

### Signal Filtering

In order to de-noise the contaminated EEG signal, filtering was applied to the input signal. A band pass filter was designed and applied to remove the noises from the signal. The band pass filter allows signals between two specific frequencies (cut-off frequencies) to pass, but that discriminates against signals of other frequencies. This filter module allows to pass signals between 1 and 70 Hz, since typical brain signal rhythms are located between this frequency frame [16].

### Blind Signal Separation

BSS, also known as blind source separation, is the separation of a set of source signals from a set of mixed signals, without the aid of information about the source signals or the mixing process. In EEG, the interference from muscle activity masks the desired signal from brain activity. BSS, however, can be used to separate these two so an accurate representation of brain activity can be achieved [17].

IVA was formulated as a general joint BSS framework to ensure that the corresponding sources extracted from different data sets are maximally dependent while the sources within each data set are independent of each other. IVA is a generalization of ICA from one to multiple data sets, and was originally designed to address the permutation problem in the frequency domain for the separation of acoustic sources [18]. That is to say, source independence within one data set and corresponding source dependence across multiple data sets are maximized simultaneously [19].

### Performance Evaluation Model

In order to quantitatively measure and evaluate the performance of our IVA method, two objective measures were adopted. The first measure is the RMSE, and the second is SNR that is often encountered in electrophysiology [20]. For each filtered signal in both realistic simulated data and real EEG datasets, we perform individually the ICA, CCA and IVA techniques while we keep track of the respectively RMSE and SNR values.

# Results

Table 1 illustrates the Root Mean Square Error (RMSE) and Signal-to-noise-ratio (SNR )value from all five realistic simulated datasets when performing ICA, CCA and IVA methods individually (Table 1). These values were calculated by the performance evaluation module.

**Table 1.** The results of the performance evaluation module on realistic simulated data

|  | Independent Vector Analysis (IVA) | | Independent Component Analysis (ICA) | | Canonical Component Analysis (CCA) | |
| --- | --- | --- | --- | --- | --- | --- |
| Realistic Simulated Dataset | Characteristic Value | | | | | |
|  | RMSE | SNR | RMSE | SNR | RMSE | SNR |
| 1 | 0.3249 | 0.0756 | 1.6600 | 0.0220 | 6.6161 | 0.0005 |
| 2 | 0.3289 | 0.1415 | 1.0521 | 0.0035 | 6.3999 | 0.0005 |
| 3 | 0.3349 | 0.1007 | 0.6702 | 0.0060 | 2.6235 | 0.0013 |
| 4 | 0.3373 | 0.2084 | 0.6469 | 0.0066 | 3.1564 | 0.0011 |
| 5 | 0.3379 | 0.1698 | 0.7139 | 0.0056 | 2.9767 | 0.0011 |

For each realistic simulated dataset, Figure 2 illustrates the input data, the signal after applying the band pass filter and the extracted, pure EEG, signal after muscle artifact rejection using the IVA method.

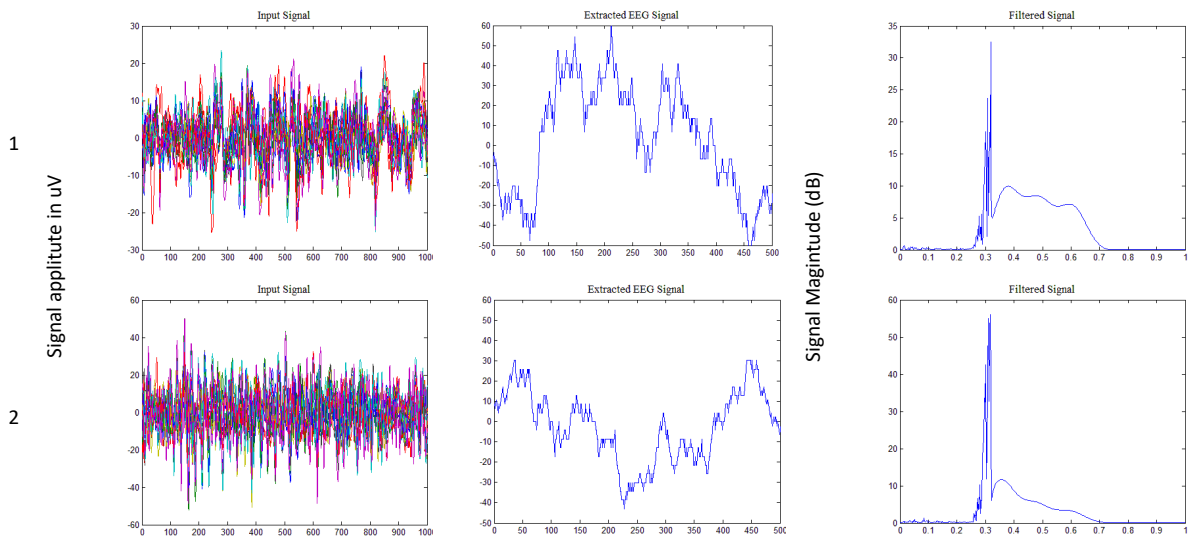

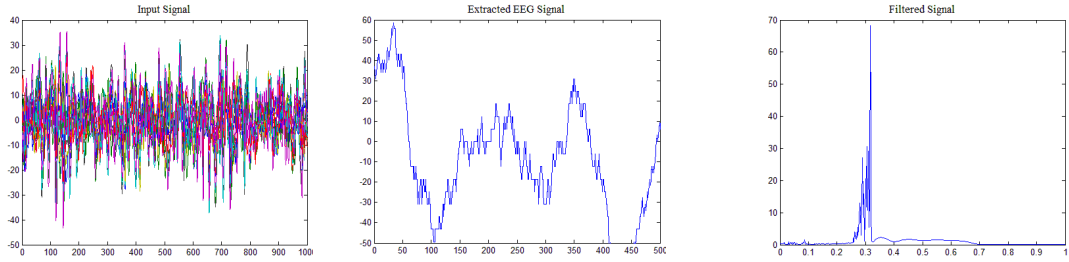
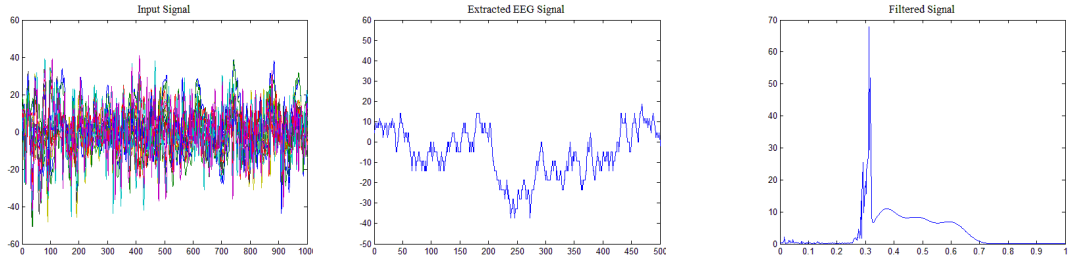
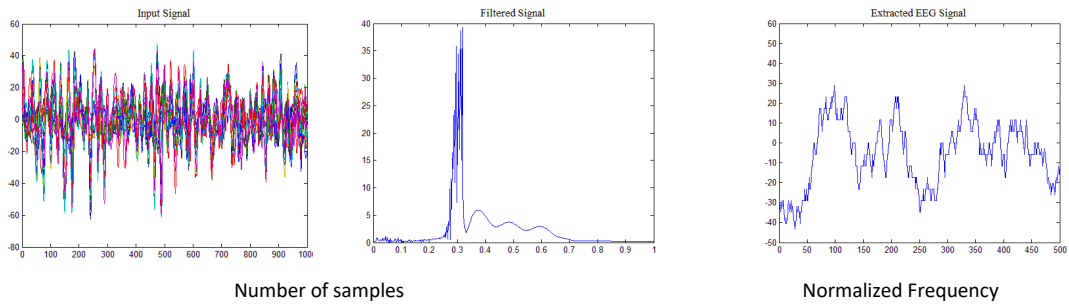

Number of samples          Normalized Frequency

**Figure 2.** The five different, realistic simulated datasets. First column presents the input signal while second column shows the signal after applying the band pass filter. Third column presents the extracted, puree EEG signal after muscle artifact rejection using the IVA method.

The proposed IVA method was also applied to a selected real EEG recording recruited from the Day Care Center of the Greek Association of Alzheimer Disease and Related Disorders "Saint Ioannis". We have selected a dataset that is annotated with a patient body movement (Figure 3).

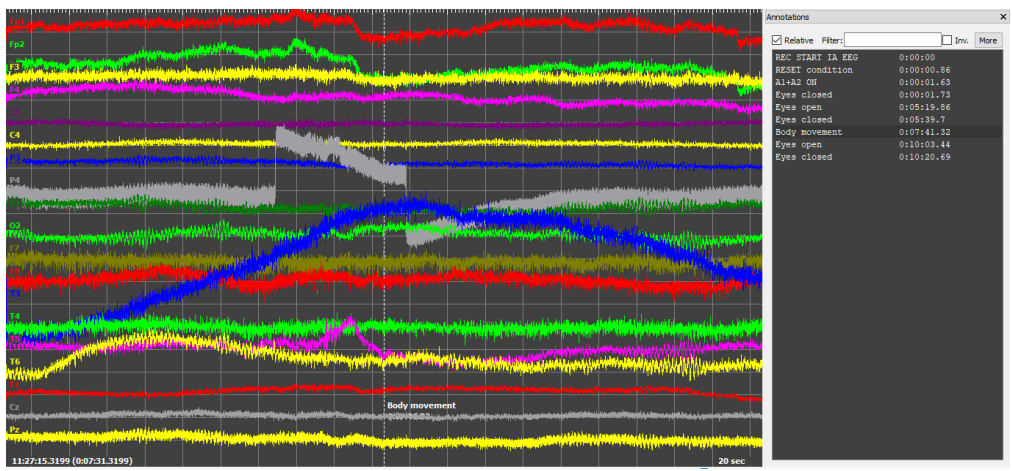

**Figure 3.** Selected real EEG data with body movement annotation. The whole recording consists almost of 10 minutes and 20 seconds while the body movement is annotated around 7 minutes and 41 seconds after recording started.

After applying to the dataset the IVA method we can extract the pure EEG signal free of muscle artifacts (Figure 4).

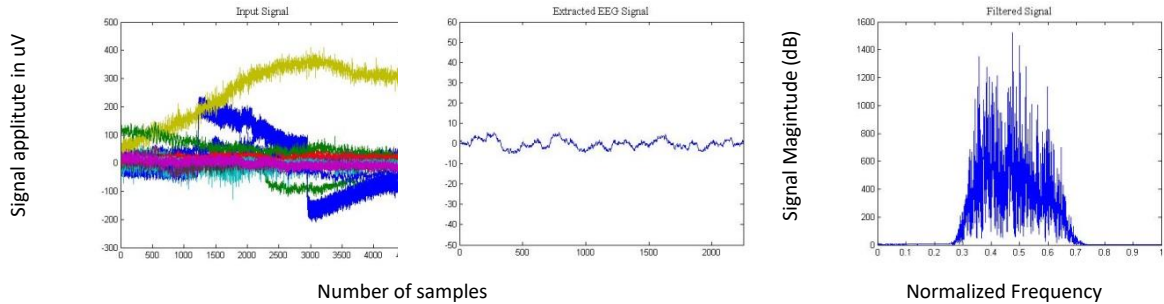

**Figure 4.** Applying the proposed IVA method into a selected, body movement annotated, EEG dataset. The figure illustrates the selected input signal, the pure EEG extracted signal and the produced signal after band pass filtering

Performance evaluation module, calculates the two objective measures that were adopted. Table 2 illustrates the RMS Error and SNR Value for the selected real EEG dataset.

*Table 2.* The results of the performance evaluation module on the selected real EEG signal

| Independent Vector Analysis (IVA) | | Independent Component Analysis (ICA) | | Canonical Component Analysis (CCA) | |
|---|---|---|---|---|---|
| Characteristic Value | | | | | |
| RMSE | SNR | RMSE | SNR | RMSE | SNR |
| 0.3333 | 0.0130 | 0.3474 | 0.0349 | 24.7643 | 0.0001 |

## Discussion

The results are in agreement with the studies that IVA method is better on suppressing muscle artifacts from EEG recordings, without removing significant underlying EEG information. This is occurred due to the fact that IVA method takes advantages of both CCA and ICA but also solves the "permutation problem" that occurred when separated data must be aligned [21] in order that the output signal must contains data from the same source.

Another advantage of the proposed system is that the latter is significant faster, in terms of time execution, in respect with classical ICA and CCA approaches. This is occurred due to fact that the diagonalization methods for IVA have been replaced, within the proposed system, with the SCHUR decomposition, a faster and more effective way in diagonalization [22]. That offers a faster IVA - BSS that can be used on time demanding applications such as Brain Machine Interfaces (BMI).

Furthermore, the proposed system modules, due to their simplicity, can be used on portable and energy efficient computational systems.

# Conclusion

In case of EEG recordings from people with Subjective Cognitive Impairment (SCI), Mild Cognitive Impairment (MCI) and dementia, the necessity of removing muscle artifacts is of the essence since these recordings must be used for further evaluation. Often, these EEGs involve patient's movement and thus muscular activity. In order to measure the functional connectivity between brain regions and activation areas during cognitive performance, these muscle artifacts must be removed. Our study indicates that the proposed IVA - BBS overcomes with this need since both objective measures of RSME and SNR are significant low and muscle artifacts are removed successfully from the original recordings.

*The authors declare that they have no conflicts of interest.*